\documentstyle[12pt,epsf]{article}

\newcommand{\bmat}{\left(\begin{array}}
\newcommand{\emat}{\end{array}\right)}
\def\NPB#1#2#3{Nucl. Phys. B{#1} (19#2) #3}
\def\PLB#1#2#3{Phys. Lett. B{#1} (19#2) #3}

\def\PRD#1#2#3{Phys. Rev. D{#1} (19#2) #3}

\def\yzero{\smash{\hbox{$y\kern-4pt\raise1pt\hbox{${}^\circ$}$}}}

\def\th{\theta}

\def\-{\hphantom{-}}
\def\ov{\overline}
\def\s2{\frac{1}{\sqrt2}}

\def\beq{\begin{equation}}
\def\eeq{\end{equation}}
\def\beqa{\begin{eqnarray}}
\def\eeqa{\end{eqnarray}}

\def\Tr{{\rm Tr \,}}

\def\IF{\relax{\rm I\kern-.18em F}}
\def\II{\relax{\rm I\kern-.18em I}}
\def\IP{\relax{\rm I\kern-.18em P}}
\def\inbar{\vrule height1.5ex width.4pt depth0pt}
\def\IC{\relax\hbox{\kern.25em$\inbar\kern-.3em{\rm C}$}}
\def\IR{\relax{\rm I\kern-.18em R}}

\def\cc{{\cal C}}
\def\ck{{\cal K}}

\def\cm{{\cal M}}

\def\cz{{\cal Z}}

\def\Dsl{\,\raise.15ex\hbox{/}\mkern-13.5mu D} 
\def\IZ{Z}

\newcommand{\drawsquare}[2]{\hbox{%
\rule{#2pt}{#1pt}\hskip-#2pt
\rule{#1pt}{#2pt}\hskip-#1pt
\rule[#1pt]{#1pt}{#2pt}}\rule[#1pt]{#2pt}{#2pt}\hskip-#2pt
\rule{#2pt}{#1pt}}

\newcommand{\fund}{\raisebox{-.5pt}{\drawsquare{6.5}{0.4}}}
\newcommand{\Yasymm}{\raisebox{-3.5pt}{\drawsquare{6.5}{0.4}}\hskip-6.9pt%
        \raisebox{3pt}{\drawsquare{6.5}{0.4}}}
\newcommand{\antifund}{\overline{\fund}}

\topmargin
-1.5cm
\textwidth
15.5cm
\textheight
24cm
\oddsidemargin
0.7cm
\evensidemargin
1.2cm

\begin{document}

\makeatletter
\@addtoreset{equation}{section}
\makeatother
\renewcommand{\theequation}{\thesection.\arabic{equation}}
\pagestyle{empty}
\rightline{FTUAM-98/16, IASSNS-HEP-98/72, IFT-UAM/CSIC-98-11}
\rightline{\tt hep-th/9808139}
\vspace{0.5cm}
\begin{center}
\LARGE{Anomalous U(1)'s in
Type I  and Type IIB \\
 D=4, N=1 string vacua\\[10mm]}
\large{
L.~E.~Ib\'a\~nez$^1$, R. Rabad\'an$^1$
and A.~M.~Uranga$^2$\\[2mm]}
\small{
$^1$ Departamento de F\'{\i}sica Te\'orica C-XI
and Instituto de F\'{\i}sica Te\'orica  C-XVI,\\[-0.3em]
Universidad Aut\'onoma de Madrid,
Cantoblanco, 28049 Madrid, Spain.\\[4mm]
$^2$ Institute for Advanced Study, Olden Lane, Princeton NJ 08540,
USA.\\[1mm]
 }
\small{\bf Abstract} \\[7mm]
\end{center}

\begin{center}
\begin{minipage}[h]{14.0cm}

{\small
We study the cancellation of $U(1)$ anomalies in
Type I and Type IIB  $D=4$, $N=1$ string vacua.
We first consider the case of compact toroidal $Z_N$
Type IIB orientifolds and then proceed to the non-compact case
of Type IIB D3 branes at orbifold and orientifold singularities.
Unlike the case of the heterotic string we find that
for each given vacuum one has generically more than one
$U(1)$ with non-vanishing triangle anomalies.
There is a generalized Green-Schwarz mechanism
by which these anomalies are cancelled. This involves
only the Ramond-Ramond scalars coming from the
twisted closed string spectrum but not those coming
from the untwisted sector.
Associated to the anomalous $U(1)$'s there are field-dependent
Fayet-Illiopoulos terms whose mass scale is fixed by
undetermined vev's of the NS-NS partners of the
relevant twisted RR fields. Thus, unlike what happens
in heterotic vacua, the masses of the anomalous $U(1)$'s
gauge bosons may be arbitrarily light.
In the case of D3 branes at singularities, appropriate
factorization of the $U(1)$'s constrains the Chan-Paton matrices beyond
the restrictions from cancellation of non-abelian anomalies. These
conditions can be translated to constraints on the T-dual Type IIB brane
box configurations. We also construct a new large family of $N=1$ chiral
gauge field theories from D3 branes at orientifold singularities, and
check its non-abelian and $U(1)$ anomalies cancel.}

\end{minipage}
\end{center}
\newpage
\setcounter{page}{1}
\pagestyle{plain}
\renewcommand{\thefootnote}{\arabic{footnote}}
\setcounter{footnote}{0}

\section{Introduction}

One of the most inspiring features of string theory is that it describes
consistent quantum theories of gravity and gauge interactions. For some
vacua of the theory, where gauge and/or gravitational anomalies are
potentially present, the claim above may be very non-trivial already at
the one-loop level. However, string theory always provides the appropriate
field content and interactions to yield an anomaly-free theory.
The paradigmatic example of such property is the cancellation of anomalies
in ten-dimensional heterotic of type I string theory, via the
Green-Schwarz mechanism \cite{gs}. The usual contributions
from fermions and the metric to gauge and gravitational anomalies are
cancelled by further counterterms generated by the exchange of
the two-form
field.

Different versions of this mechanism play a key role also in
compactifications of string theory to lower dimensions. The study of
its precise form and its consequences in vacua with different numbers of
spacetime dimensions and supersymmetries is an interesting subject.

For phenomenological reasons, most of the interest has centered in
the study of $D=4$, $N=1$ compactifications of the $SO(32)$ and
$E_8\times E_8$ heterotic superstrings. Perturbatively, here the pattern
of anomaly
cancellation is quite restricted \footnote{As we comment on in our final
remarks, this is also the pattern for type I compactifications on
{\em smooth} Calabi-Yau threefolds.}. At most {\em one} $U(1)$ gauge factor is
allowed to have triangle anomalies. This presents mixed gravitational
and gauge anomalies, but they are
precisely on the ratios adequate to allow for their cancellation through
the exchange of the model-independent pseudoscalar partner of the
dilaton \cite{anouno, nignos, sin}.
This is a four-dimensional version of the GS mechanism
mentioned above. The pseudo-anomalous $U(1)$ finally gets a large mass
(slightly lower than the string scale)  due to the presence of a
Fayet-Illiopoulos D-term which triggers Higgs breaking to a one-loop
stable vacuum \cite{anouno,fi}.
This beautiful mechanism is quite model independent, and has allowed to
draw a number of phenomenological interesting consequences  valid for generic
compactifications of this type (see e.g. \cite{sin,ir,ramond}
and references therein).

For $D=4$, $N=1$ compactifications of type I string theory, on the other
hand, there has not been an analogous study, even though the issue is
of similar phenomenological interest
\footnote{Refs. \cite{mr,radu,lpst} have recently appeared concerning
anomalous $U(1)$'s in the M-theory/Type I settings but their
contents have no overlap with the present work.}. The purpose of the
present paper is
to improve the understanding of anomalous $U(1)$'s in such
compactifications. We will first center on type IIB toroidal orientifolds
\cite{bl,ang,kak1,kak2,kak3,zwart,odri,afiv,lapolla,kakn,lykken,gepner} ,
since they are simple constructions whose world-sheet formulation is well
understood, and they are expected to illustrate generic properties of
type I compactifications.

We will be interested in determining the new features present in anomaly
cancellation in type I vacua, as compared with perturbative
heterotic vacua. A main novelty in type I compactifications,
as compared with
heterotic ones, is the presence of D-branes. The compact models we study
contain D9 branes and possibly one set of D5 branes. On general grounds,
one then expects a more complicated pattern of anomaly cancellation.
Indeed, as we will
discuss, these models have generically several anomalous
$U(1)$'s. Their triangle anomalies will be cancelled again by a
four-dimensional generalized
version \footnote{ An analogous mechanism for $D=6$ theories was first
considered in ref.\cite{sagcan}.} of the Green-Schwarz mechanism, but it
will involve the exchange of RR twisted fields. In particular, untwisted
fields like the partner of the dilaton do not take part in the
cancellation of anomalies. Another marked difference with respect to the
heterotic
case is that the Fayet-Illiopoulos parameters are controlled by
twisted fields, the NS-NS partners of the RR fields mentioned above. This
allows one to tune the FI to any desired scale by merely tuning the vevs
of these blowup modes. The $U(1)$'s are spontaneously broken, but their
masses are not tied up to the string scale, and can be very light.

Thus, type I compactifications allow to understand the physics of
cancellation of anomalies in the presence of D branes. This is an
interesting point, since type I models with D5 branes are dual to
heterotic compactifications with NS fivebranes. These last vacua are
highly non-perturbative and the analysis of the dual type I can provide
some insight into their properties. However, not much is known about their
explicit construction, especially in four dimensions, and so this
discussion is beyond the scope of the present paper.

Another application of the understanding of type I vacua with D-branes is
the study of anomaly cancellation in the world-volume of decoupled
D-branes. The
interest is that, in the decoupling limit  where gravity and other bulk
modes are switched off, the theory reduces to a supersymmetric field
theory. Thus string theory constructions can be used to learn about
quantum field theory. For instance, in the context of $D=6$, $N=1$ field
theories, the
study of type IIB and type I D5 branes at singularities have provided a
construction of large families of interacting superconformal field
theories
\cite{dm,intri,blumintri}. A non-trivial check of the consistency of
this construction is
that the underlying string theory ensures the cancellation of anomalies
in the world-volume field theory. In this respect, a key role is played by
the cancellation of $U(1)$ anomalies through a six-dimensional version of
the GS mechanism first uncovered in the study of compact
$D=6$ type I models \cite{sagcan,gsw,blpssw}.

Recently, some configurations of branes in string theory have been
proposed for the study of four-dimensional gauge theories. In the most
interesting case of $N=1$ supersymmety, the construction of large families
of chiral theories has been accomplished by the so-called brane box
models, consisting of grids of two kinds of NS fivebranes in type IIB
string theory, on which D5 branes are suspended \cite{hz}.
By applying T-duality to some of these models \cite{hu}, these field
theories are realized in the world-volume of D3 branes on a threefold
singularity \cite{dgm,lnv}. This construction has become very popular,
since, by use of the AdS/CFT correspondence \cite{ads}, it allows a very
simple construction of $N=1$ superconformal field theories (in the
limit of large number of D3 branes) \cite{ks,lnv,bkv}.

These configurations of D3 branes at singularities are
four-dimensional analogs of the six-dimensional theories
mentioned above. So they can also be analyzed by perturbative type I
string theory, and in particular it is a natural question to address how
their cancellation of $U(1)$ anomalies occurs. We will fist analyze the
system of D3 branes on top of orbifold  singularities, and show explicitly
how the string consistency conditions ensure the cancellation of $U(1)$
anomalies via the GS mechanism uncovered in compact models. Moreover, the
generation of FI terms gives masses to these
$U(1)$'s and explains why they are not present in the low energy dynamics.
This `freezing' of the $U(1)$'s was expected from the brane box point of
view, in analogy with the result in \cite{wit4d} for $N=2$ theories,
but had not been shown explicitly. Here we show it in the T-dual version.

Finally, we consider a family of new $N=1$ chiral gauge field theories
obtained from D3 branes on top of threefold orientifold singularities.
We show that for each singularity of a certain type one can build a theory
that becomes superconformal in the large $N$ limit. The
stringy construction of this large family of models ensures their
consistency, and we illustrate this by checking explicitly the
cancellation of non-abelian and $U(1)$ anomalies.

\medskip

The paper is organized as follows. In Section~2 we introduce the general
idea for the GS mechanism in type I and type IIB $D=4$, $N=1$
vacua. In Section~3 we center on the case of compact toroidal
orientifolds, where we show in detail the cancellation of gravitational,
mixed gauge and cubic $U(1)$ anomalies in several non-trivial examples. In
Section~4 we describe the construction of the field theories on the
worldvolume of D3 branes on top
of orbifold and orientifold singularities. We give general expressions for
the triangle anomalies, and show explicitly how the consistency conditions
for the underlying string theory ensure their cancellation via the GS
mechanism. In Section~5 we comment on the generation of FI terms, and
compare the situation with that in (perturbative) heterotic vacua. In
Section~6 we make some final remarks. Details on the construction of the
theory of D3 branes at orientifold singularities are left for the
appendix.

\section{Anomalous $U(1)$'s in Type I  and Type IIB $D=4$, $N=1$ vacua}

In what follows we are going to discuss four-dimensional vacua with
$N=1$ based on orientifolds \cite{sagnotti,hor,bs}
and/or orbifolds \cite{dhvw} of Type IIB theory.
A first class of theories will consist of compact models
from toroidal IIB orientifolds. Due to compactness the structure,
type and number of D-branes present in the vacuum will be strongly constrained
by tadpole cancellation conditions. We will concentrate here in models
which contain nine-branes and at most one set of 32 fivebranes.
The second class of theories we will be interested on will be
Type IIB brane-box models in which appropiate configurations
of D-fivebranes and NS-fivebranes are situated in such a way that
one has an effective $D=4$, $N=1$ chiral theory on the worldvolume
of the D-branes. These are non-compact theories which can also
be studied by going to their T-dual which consists on sets
of D3-branes sitting on $Z_N$ type singularities. Being non-compact,
tadpole cancellation conditions are much milder for these theories
and, e.g., the overall number of D-branes is undetermined.

In any of the above class of theories one obtains a general gauge group of the
form
\beq
\prod _{\alpha } \ (
\prod _{i=1}^{n_{\alpha }}U(1)_i\times \prod_{j=1}^{m_{\alpha }}G_j )
\eeq
where $G_j$ are  general non-Abelian groups.
Here $\alpha $ runs over the different sets of D-branes present in each model
and $n_{\alpha}$($m_{\alpha}$) are the number of $U(1)$'s (non-Abelian
groups) present in the D-brane sector $\alpha$.
Thus, for example, $\alpha $ runs over one set of 9-branes and one set of
5-branes for the compact orientifold models discussed bellow.
In general
we will have one $U(1)$ factor for each $SU(n)$ factor in the theory.

We are interested in the cancellation of $U(1)$ anomalies in these classes of
theories. There will be mixed $U(1)_i\times G_j^2$ anomalies as well
as cubic $U(1)_i\times U(1)_l^2$ anomalies. In addition, in the case of
compact orientifolds we will have to care also about mixed $U(1)$-gravitational
anomalies. The relevant graphs contributing to these anomalies are depicted in
Figure 1, for the particular case of mixed $U(1)$  non-abelian anomalies.

\begin{figure}
\centering
\epsfxsize=5.5in
\hspace*{0in}\vspace*{.2in}
\epsffile{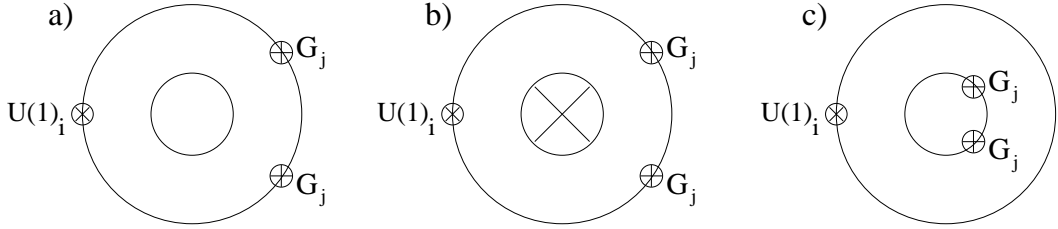}
\caption{\small The three graphs contributing to the mixed
$U(1)_i \times G_j^2$ mixed anomaly. The first two, the annulus (a) and
Moebius strip (b), give
the usual triangle anomaly of field theory. The last diagram (c), the
non-planar annulus, cancels the contribution from the first two.}
\label{fig:diagramas}
\end{figure}

The first graph corresponds to the annulus contribution
and the second to the Moebius strip (the latter is only pressent in
an orientifold setting). These two contributions correspond to the
usual effective low energy field theory triangle diagram calculation and we will
not discuss it here any further. It is the third graph in Fig.1 which will be
in charge of the cancellation of the $U(1)$ anomalies left over by the naive
triangle graph calculation. Notice that this third graph is only relevant
for anomalies involving at least one $U(1)$. Fig.2 shows the same graph
but in the closed string exchange chanel. It shows the field
theory meaning of the cancellation mechanism. A $U(1)$ couples to some
(Ramond-Ramond) closed string state which propagates and finally couples
to a pair of gauge bosons (or gravitons).

\begin{figure}
\centering
\epsfxsize=3in
\hspace*{0in}\vspace*{.2in}
\epsffile{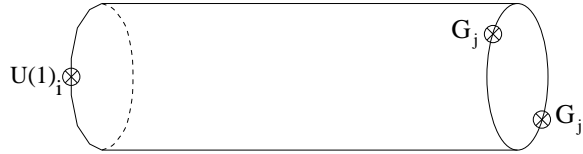}
\caption{\small The annulus diagram of Figure~1c in the
closed string channel. Twisted RR fields propagating along the
cylinder provide the contribution that cancels the anomaly.}
\label{fig:closed}
\end{figure}

The coupling of the U(1) to
a RR field described by an antisymmetric field $B_k^{\mu \nu }$ gives
rise to an effective term in the low energy action proportional to
\beq
Tr(\gamma _k \lambda _i )  B_k\wedge F_{U(1)_i}
\label{contrat}
\eeq
 where $\lambda _i$ is the CP matrix asociated to the $U(1)_i$ generator
and $\gamma _k$ is the  matrix associated to the $\theta ^k$ twist.
Notice that for RR fields belonging to the untwisted sector
this term is proportional to $Tr\lambda _i$ which vanishes for traceless
CP generators, which is the case of the compact orientifolds discussed
below.
The coupling to the right hand side of Fig.~2 gives rise to
effective low energy couplings proportional to
\beq
Tr(\gamma _k ^{-1} \lambda ^2_G) (\partial^{[\mu }B^{\nu \rho ]}_k) W^{CS}_{\mu \nu \rho }
\label{trazados}
\eeq
where $\lambda _G$ is the CP matrix associated to the rightmost gauge bosons.
$W^{CS}_{\mu \nu \rho }$ is the Chern-Simons (CS) tensor of those gauge fields.
In the case of mixed gravitational anomalies one replaces the gauge CS
tensor
by the Lorentz one and $\lambda _G$ by the unit matrix. Notice that
sectors $k$ with $Tr\gamma _k=0$ will thus not contribute to the
cancellation of mixed gravitational anomalies.

The full low energy amplitude contributing to the mixed $U(1)_i\times
G^2_j$ anomaly coming from graph 2 will be proportional to

\beq
A_{ij}^{\alpha \beta}\ = {1\over {|P|}} \sum _k A_{ij}^{\alpha \beta }(k)\ =
\
{i\over |P|}\ \ \sum_{k\in sectors} \
C_k^{\alpha \beta }\
Tr(\gamma_{\theta ^k}^{\alpha } \lambda _i^{\alpha } )\
Tr( (\gamma_{\theta ^k}^{\beta })^{-1} (\lambda_{j}^{\beta})^2 )
\label{master}
\eeq

where $|P|$ is the order of the orientifold or
orbifold group and the sum
runs over the different  twisted sectors.
The indices $\alpha (\beta )$ indicate the brane sector
from which the $U(1)_i (G_j)$ are coming from.
The coefficients $C_k$ depend on the particular twist $k$.
For $Z_N$ they are given by
\beq
 C_k^{\alpha \beta } \ = \
\prod _{a=1}^3 2\sin\pi kv_a
\label{generalck}
\eeq
where the product extends only over complex planes $a$
with $NN$ or $DD$ boundary conditions.
Here $v=(v_1,v_2,v_3)$ is the compact space twist vector,
i.e., the $Z_N$ twist is generated by rotations by
$\exp(2i\pi v_a)$ in the three $a=1,2,3$ compact complex
dimensions. The extension to $Z_N\times Z_M$ twists is straightforward.
The existence of these $C_k$ factors is crucial in obtaining
anomaly cancellation. Their existence
can be inferred from the
cylinder tadpole amplitudes shown in
the appendix of ref.\cite{afiv}. We will also  show later
how they appear naturally in the context of D-branes sitting
at orbifold/orientifold $Z_N$ singularities coming from
a discrete Fourier transform.

The $U(1)$'s which turn out to be anomalous may be written as
linear combinations of the form :
\beq
Q_k(\beta ,p) \ =\ \sum _i \sum _{ \alpha } A_{ip}^{\alpha \beta }
(k) \ Q_i^{\alpha }
\label{combi}
\eeq
For each $k$ one gets  a number of linear combinations labeled by $\beta$
and $p$. Only a subset of them are in general linearly independent.
 The number of anomalous
$U(1)$'s in a given model will depend on the number of linearly
independent
$Q_k(\beta , p)$  generators one finds. Examples will be given below.

\section{Anomalous $U(1)$s in compact Type IIB $D=4$, $N=1$ orientifolds}

As a first application of the above discussion we are going to study
cancellation of $U(1)$ anomalies in a class of $D=4$, $N=1$ Type IIB
toroidal orientifolds
(for details about these models and notation see \cite{afiv})
. We will consider models obtained by
compactifying Type IIB theory on orbifolds $T^6/P$, $P$ being
either $Z_N$ or $Z_N\times Z_M$. In this compact case these discrete
symmetries are restricted to act crystalographically on $T^6$
which substantially reduces the possibilities. Those were clasified in
\cite{dhvw,fiq}. In addition we are going to twist the theory by
the world-sheet parity operator $\Omega $. As a result one gets
$D=4$ theories with $N=1$ unbroken supersymmetry. Orientifolds with
only odd order $Z_N$ twists have only 9-branes. Those with an
order-two twist (acting e.g. along the first two $a=1,2$ complex
coordinates) will have in addition one set of 5-branes with their
worldvolume filling the  four non-compact dimensions plus the third
complex plane. For simplicity we will restrict ourselves to the
case of orientifolds with only one twisted sector of order two.
These will have only one set of 5-branes.

 Unlike the non-compact case
the total number of each type of D-brane in the vacuum is limited due to
tadpole cancellation conditions. The cases with maximal symmetry will admit
a maximum of 32 9-branes and 32 5-branes. In their worldvolumes will live
gauge fields with associated CP matrices $\lambda ^{\alpha }$, $\alpha =9,5$.
Those will be $32\times 32$ hermitian matrices. As explained in
\cite{afiv}
in this class of orientifolds with an $\Omega $ action it is usefull to use a
Weyl-Cartan realization of the CP algebra. The 9-brane and 5-brane CP matrices
in this case are restricted to be $SO(32)$ generators.
 They can be organized
into charged generators $\lambda_a = E_a$, $a=1,\cdots, 480$, and Cartan
generators $\lambda_I = H_I$, $I=1,\cdots, 16$.  The twist
matrices $\gamma_{\theta }^{\alpha }$ and its powers represent
 the action of the e.g. $Z_N$ group
on Chan Paton factors, and they correspond to elements of a  discrete
subgroup of the Abelian group spanned by the Cartan generators.
Hence, we can write
\begin{equation}
\gamma _{k }^{\alpha } \equiv \gamma _{\theta ^k }^{\alpha }= e^{-2i\pi k
V_{\alpha }
\cdot H }
\label{Vdef}
\end{equation}
This equation defines the 16-dimensional `shift' vector $V_{\alpha }$.
One can see that $\gamma _1^{\alpha } $ can be chosen
diagonal and furthermore $(\gamma_1^{\alpha} )^N = \pm 1$.
 Cartan generators $H_I$, are represented by tensor products of
$2 \times 2$ $\sigma_3$ submatrices. We chose the normalization
of the $SO(32)$ generators $\lambda $ in such a way that
$Tr\lambda ^2 =1$. The (unnormalized)
 generator $\lambda _i$ of a given $U(1)_i$ will be given by
a linear combination
\beq
 \lambda _i= \sum _{\alpha
=9, 5 } Q_i^{\alpha }\cdot H^{\alpha }
\eeq
 where $Q_i$ is a 16-dimensional
real vector. These $U(1)_i$ generators commute with the
generators of the non-Abelian groups. In this class of models, in the absence
of continuous Wilson lines and/or if all 5-branes sit at fixed points, there are
only unitary or orthogonal groups (otherwise symplectic factors do also
appear).
There is a $U(1)$ factor for each of the $SU(n)$'s in the model. Thus
the 16-dimensional
vectors $Q_i^{\alpha }$ will have typically the form
$Q_i^{\alpha }=(0,0,.,0,1,..,1,0,0,..,0)$ where the  n one-entries sit at the possition
where a corresponding $SU(n-1)$ lives. With these conventions and normalizations
the relevant traces for anomaly cancellation are:
\beq
\Tr(\gamma _k^{\alpha }\lambda _i^{\alpha })=
\Tr(e^{-2i\pi kV^{\alpha }\cdot H }\ Q_i^{\alpha }\cdot H^{\alpha })=
(-i)\ 2n_i\sin2\pi kV^{\alpha }_i
\eeq
where $n_i$ is the rank of the $U(n)$ group containg this $U(1)$ and
$V^{\alpha }_i$ is the component of the $V^{\alpha }$ vector along any of the
entries overlapping with that $U(n)$.
Notice that the $n_i$ factor appears because we
have not normalized the $U(1)$. In the same way one obtains:
\beq
\Tr((\gamma _k^{\beta })^{-1}(\lambda _j^{\beta })^2)=
\Tr(e^{2i\pi kV^{\beta }\cdot H }\  (\lambda _j^{\beta })^2  )= \cos2\pi
kV^{\beta
}_j
\eeq
where again $V^{\alpha }_j$ is the component of the shift vector along any
entry overlapping with the group $G_j$. Thus for the present class of
compact orientifolds with  $|P|=2N$
one gets a total contribution to the mixed anomalies from
the graph in fig.2
\beq
A_{ij}^{\alpha \beta}\ =\
{1\over N}\ \ \sum_{k\in sectors} \
C_k^{\alpha \beta }(v)\
n_i\sin2\pi kV^{\alpha }_i \
\cos2\pi kV^{\beta }_j
\label{masterorient}
\eeq
where $k$ runs over twisted $Z_N$ sectors, $\alpha , \beta $ run
over 5,9 (meaning 5- or 9-brane origin of the gauge boson).
The coefficients are given by
\beqa
 C_k^{\alpha \beta }(V) \ = \
&\prod _{a=1}^3 2\sin\pi kv_a  \ \ \ ; \ &\alpha = \beta \\
=\ & 2\sin\pi kv_3 \ \ \ ;\ \ &\alpha \not= \beta \nonumber
\eeqa
Notice that the case $\alpha \not= \beta $ corresponds to mixed anomalies
mixing gauge groups coming from different brane systems (9-branes and
5-branes).

\medskip

In the case of cubic $U(1)$ anomalies a similar formula is obtained.
The only difference is that, since we have not normalized the  $U(1)_j$
generators there will be an extra factor $(2n_j)$ in this expression.
Care must be taken also to recall that there is an extra
symmetry factor $1/3$
when computing the triangle graphs for $U(1)_i^3$ compared to
those for $U(1)_i\times U(1)_j^2$, $i\not=j $. Altogether one  has
an expression for the mixed $U(1)_i\times U(1)_j^2$ anomalies:
\beq
A_{ij}^{\alpha \beta}\ =\
{2\over N} \ \sum_{k\in sectors} \
C_k^{\alpha \beta }(v)\
n_i\sin2\pi kV^{\alpha }_i \
n_j\cos2\pi kV^{\beta }_j
\label{masterunos}
\eeq
In the case of mixed $U(1)$ gravitational  anomalies one has
\beq
A_{i}^{\alpha }\ =\
 {3\over 4} {1\over N}\ \ \sum _{\beta }  \sum_{k\in sectors} \
C_k^{\alpha \beta }(v)\
n_i\sin2\pi kV^{\alpha }_i \
\Tr((\gamma _k^{\beta } )^{-1})
\label{mastergrav}
\eeq
where ${3\over 4}={{24}\over {32}}$ is a normalization factor.
 From the above expressions an interesting sum rule relating the
mixed $U(1)$-gauge anomalies and the $U(1)$-gravitational anomalies
can be obtained:
\beq
A_i^{\alpha }\ =\ {3\over 2} \sum_j \sum_{\beta }\  n_j\ A_{ij}^{\alpha \beta }
\label{sumrule}
\eeq
where $n_j$ is the rank of the $j^{th}$ $U(n)$ or $SO(m)$ gauge group in
the theory and the $j$-sum goes over all of them.
This formula is interesting because it is model independent, i.e.,
one only needs to know the massless spectrum of the theory
and the $U(1)$ charges in order to compute it, without any reference
to the specific form of Chan-Paton matrices nor twist structure.

Notice the following points in these expressions:

i) Untwisted RR fields, like the partners of the dilaton $Re\,S$ and
the partners of the untwisted moduli $Re\,T_b$ do not participate in
anomaly cancellation since in this case $\gamma _k =1$ and
$\Tr\lambda _i=0$.

ii) Due to the $C_k$ coefficients, RR twisted fields associated to
twists leaving one torus fixed do not contribute to anomaly
cancellation for $\alpha =\beta $. On the other hand they do in general
contribute to cancellation of anomalies mixing different types of
branes ($\alpha \not= \beta $).

iii) Twisted sectors with $\Tr\gamma _k = 0$ do not contribute to
the cancellation of the mixed $U(1)$-gravitational anomalies.

Up to now  we have considered the most symmetric situation in which
Wilson lines are absent and all 5-branes sit at the fixed point at the origin.
Something analogous happens in more general cases. Consider for example
the case of an orientifold with a quantized Wilson line background.
As explained in \cite{afiv} , now the different fixed points of
the orbifold will not be all equivalent since different fixed points
will have associated different $\gamma _k^{(9)}$ matrices.
In particular
the orbifold action  is
generated by the space group which involves elements $(\theta ,1)$, with
$\theta $  representing $Z_N$ rotations, and elements
$(1,e_m)$, with $e_m \in \Lambda$, $m=1,\cdots, 6$, where $T^6=R^6/\Lambda$.
 The element $(\theta ^k,1)$
is embedded in the open string sector through unitary
matrices $\gamma _k^{\alpha }$.
In addition there can be background Wilson lines
which correspond to embeddings of the elements $(1,e_m)$ through matrices $W_m$
into the 9-brane sector. To a fixed point $f$ of $\th^k$
there corresponds an element $(\theta^k, c_m e_m)$ such that
$(1-\th^k)f = c_m e_m$, for some integers $c_m$.
The 9-brane monodromy associated to this
fixed point $f$ will thus be $\gamma_k^f=(\prod_m W_m^{c_m})\gamma_k ^{9}$.
Thus different fixed points f will have different
twist matrices acting on the CP factors. Thus in this case we will have
\beq
A_{ij}^{(99)}\ =\
{1\over N}{1\over {N_f}}\ \ \sum _f^{N_f} \sum_{k\in sectors} \
C_k^{(99) }(v)\
n_i\sin2\pi kV^{f }_i \
\cos2\pi kV^{f }_j
\eeq
where the additional sum goes over the different fixed points which
feel different monodromy $\gamma _k^f=\exp(-i2\pi kV^f \cdot H)$.
An example is provided below.

\subsection{Examples}

Let us see how the $U(1)$ anomaly cancellation proceeds in some specific
orientifold examples.

{\it i) The $Z_3$ $D=4$, $N=1$ orientifold}

This is perhaps the simplest compact orientifold in four
dimensions
\cite{ang,kak1}. It is obtained
by modding Type IIB theory on a torus $T^6$ by the standard $Z_3$
action with $v={1 \over 3}(1,1,-2)$. In this case there are only
9-branes and tadpole cancellation conditions require
$\Tr\gamma _{\theta }=\Tr\gamma _1=-4$. The unique solution (up to
irelevant phases) is
$\gamma _1= \exp(-2i\pi V\cdot H)$ with a shift \cite{afiv}
\beq
V\ ={1\over 3} (1,1,1,1,1,1,1,1,1,1,1,1,0,0,0,0)
\eeq
The gauge group is $U(12)\times SO(8)$ and the charged chiral matter fields
transform as $3(12, 8_v)_{1}+3({\overline {66}},1)_{-2}$, where the subindex
shows the $U(1)$ charges. With the standard normalization for generators of
non-Abelian groups one finds mixed $U(1)$ triangle anomalies with respect to
$SU(12), SO(8)$ and the $U(1)$ equal to $-18, 36$ and $-432$
respectively. Let us now compute the contribution of eq.
(\ref{masterorient}) to these anomalies.
Now we have only one constant $C^{\alpha \beta }_1=$$-C^{\alpha \beta }_2=-3\sqrt{3}$.
On the other hand $V_i=1/3$, $V_{SU(12)}=1/3$ and $V_{SO(8)}=0$. Thus altogether
we get
\beqa
A_{U(1)}(SU(12), SO(8), U(1))
&=& {1\over 3}2(-3\sqrt{3} )\ (12\ {{\sqrt{3}}\over 2})\times (-{1\over 2}, 1, 24(-{1\over 2}) )
\nonumber \\
&=&(18, -36, 432)
\eeqa
where the factor $2$ comes from the sum over $k=1,2$. This contribution exactly cancels
that from the triangle graphs. Concerning the $U(1)$-gravitational anomaly
the triangle graph gives a contribution proportional to $(-108)$ whereas eq.
(\ref{mastergrav}) yields
\beq
A_{grav}\ =\ {{3}\over 4}2(-3\sqrt{3})(6\sqrt{3})(-4) \ =\ 108
\eeq
as should be.

\medskip

{\it ii) The $Z_3$ orientifold with Wilson lines}

The above example was quite simple since it has only one anomalous $U(1)$.
Things get a little bit more involved when therere are several $U(1)$s.
Let us consider an orientifold obtained from the previous one by the addition
of a discrete Wilson line asociated to a shift
$W=1/3(0,0,1,1,2,2,$$0,0,1,1,2,2,1,1,1,1)$
 e.g., around the first complex dimension \cite{afiv,fin} .
 Then the 27 fixed points of the orbifold are split into
three sets of nine each which have asociated CP twists respectively :
\beqa
V\ &=\ {1\over 3}  (1,1,1,1,1,1,1,1,1,1,1,1,0,0,0,0) \nonumber \\
V+W\ &=\ {1\over 3} (1,1,2,2,0,0,1,1,2,2,0,0,1,1,1,1) \nonumber \\
V-W\ &=\ {1\over 3} (1,1,0,0,2,2,1,1,0,0,2,2,2,2,2,2)
\eeqa
The gauge symmetry is $U(4)^4$ and
the charged particle spectrum consists of chiral fields
$3(1,{\bar 4},4,1)_{(0,-1,-1,0)}$$
+3(1,4,1,{\bar 4})_{(0,1,0,-1)}+$$
3(1,1,{\bar 4},4)_{(0,0,1,1)}+$$
3(6,1,1,1)_{(-2,0,0,0)}$.
Here the four subindices correspod to the $U(1)$ charges. We label the $U(1)$'s and the
$SU(4)$'s respectively with the indices $i=1,\ldots,4$ and $j=1,\ldots,4$.
Computation of the $U(1)_i\times SU(4)^2_j$ mixed anomalies yields the matrix
$$A_{ij}^0\ =\ \pmatrix{
-6 & 0 & 0 & 0 \cr
0  & 0 & -6 & -6 \cr
0  & -6 & 0 & 6  \cr
0  & 6  &  6 & 0  \cr }
$$
Let us now compute the contribution coming from eq.(\ref{masterorient})
\beqa
A_{ij}\ = \  {1\over 3}{1\over 3}2\ (-3\sqrt{3} )
& \left\{  (-\sqrt{3} )  \pmatrix{
1 & 1& 1& 0\cr
1 & 1& 1 & 0\cr
1 & 1 & 1 & 0 \cr
-2 & -2 & -2 & 0 \cr
} \  +\  (-\sqrt{3} )
 \pmatrix{
1 & -1& 0 & 1\cr
1 & -1& 0 & 1\cr
-2 & 2 & 0 & -2 \cr
1 & -1 & 0 & 1 \cr
}
\ +\ \right. \nonumber \\
& + \  \left. (-\sqrt{3} ) \pmatrix{
1 & 0& -1& -1\cr
-2 & 0& 2 & 2\cr
1 & 0 & -1 & -1 \cr
1 & 0 & -1 & -1 \cr
}
 \right\} \ =\  -\ A_{ij}^0
\eeqa
so that mixed anomalies are exactly cancelled. In the present case
three out of the four $U(1)$s are anomalous. It is easy to check that
cubic anomalies do also cancel in a similar fashion. Concerning the mixed gravitational
anomalies, triangle graphs yield $(-36,0,0,0)$. The sum rule
(\ref{sumrule}) leads to
${3\over 2}\ 4 (6, 0, 0, 0) $ which exactly cancels that contribution.

\medskip

{\it   iii) An example with both 9-branes and 5-branes: the $Z_6$' orientifold}

In this example \cite{zwart,afiv}
the twist $\theta $ is generated by $v={1\over 6}(1,-3,2)$.
There is an order-two twist corresponding to $3v$ and tadpole cancellation
conditions require the presence of 32 9-branes and 32 5-branes.
Here we will consider a configuration with all 5-branes sitting at the
fixed point at the origin.
The tadpole cancellation conditions may be found in \cite{zwart,afiv}. The
corresponding twist on CP matrices is generated by
$\gamma _1^{\alpha }=\exp(-i2\pi V^{\alpha }\cdot H)$ with \cite{afiv}
\beq
V^9\ =\ V^5\ =\ {1\over {12}}(1,1,1,1,5,5,5,5,3,3,3,3,3,3,3,3).
\eeq
The gauge group is $(U(4)\times U(4)\times U(8))^2$ and the spectrum may be
found in table 3 of \cite{afiv} . This model has three $U(1)_i^9$,
$i=1,2,3$, from the 9-brane sector and three $U(1)_i^5$, $i=1,2,3$,
from the 5-brane sector. Their mixed anomalies with respect to the six
non-Abelian groups are found to be:
$$
A^{\alpha \beta }_{ij}
 \ =\
\pmatrix{
2 & 2 & 8 & -2 & 0 & -4 \cr
-2 & -2 & -8 & 0 & 2 & 4 \cr
0 & 0 & 0 & 2 & -2 & 0 \cr
-2 & 0 & -4 & 2 & 2 & 8 \cr
0 & 2 & 4 & -2 & -2 & -8 \cr
2 & -2 & 0 & 0 & 0 & 0 \cr
}
$$
were the two $3\times 3$ sub-matrices in the diagonal correspond to
anomalies not
mixing fields from 9-branes to those from 5-branes. The off-diagonal boxes
correspond to the contribution from particles being charged under both
9-brane and 5-brane gauge groups. The columns label the $U(1)$'s
whereas the rows label the $SU(N)$'s.
Let us now see how these anomalies are cancelled by the exchange of RR-fields.
The $C_k^{\alpha \beta }$ factors play now an important role. For the
present case one finds:
\beqa
C_1^{99} &= &C_1^{55}=-C_5^{99}=-C_5^{55}=\ -2\sqrt{3} \nonumber \\
C_2^{99}&=&C_2^{55}=C_4^{99}=C_4^{55}\ =\ 0 \nonumber \\
C_3^{99} &=& C_3^{55}=C_3^{95}\ =\ 0 \nonumber \\
C_1^{95}& =& C_2^{95}=-C_4^{95}=-C_5^{95}\ =\ \sqrt{3}\ . \nonumber \\
\eeqa
 From this we conclude that twisted sector $\theta ^3$ does not
 contribute to the cancellation. Twisted fields from sector $k$ and
$N-k$ yield the same contribution thus altogether one gets twice the sum of two
contributions from $k=1,2$ sectors:
\beq
-{2\over 6} (\sqrt{3})^2\left[
\pmatrix{ 2 & 2 & 8 & -1 & -1 & -4 \cr
-2 & -2 & -8 & 1 & 1 & 4 \cr
0 & 0 & 0 & 0 & 0 & 0 \cr
-1 & -1 & -4 & 2 & 2 & 8 \cr
1 & 1 & 4 & -2 & -2 & -8 \cr
0 & 0 & 0 & 0 & 0 & 0 \cr
}
 +
\pmatrix{
0 & 0 & 0 & -1 & 1 & 0 \cr
0 & 0 & 0 & -1 & 1 & 0 \cr
0 & 0 & 0 & 2 & -2 & 0 \cr
-1 & 1 & 0 & 0 & 0 & 0 \cr
-1 & 1 & 0 & 0 & 0 & 0 \cr
2 & -2 & 0 & 0 & 0 & 0 \cr
}
\right]
\label{matrices}
\eeq
which exactly cancels the triangle anomalies. Notice that
the $\theta ^2 (\theta ^4 )$ sector only contributes to the
cancelation of mixed anomalies of 9(5)-brane $U(1)$'s with
5(9)-brane gauge groups. One can check that out of the
six $U(1)$'s one can form two linear combinations which are
anomaly free. Thus this orientifold has four anomalous $U(1)$s.
Equation (\ref{combi}) allows us to obtain which four
linear combinations are anomalous:
\beqa
&  & 2Q_1^9+2Q_2^9+8Q_3^9-Q_1^5-Q_2^5-4Q_3^5 \nonumber \\
&   & -Q_1^9-Q_2^9-4Q_3^9+2Q_1^5+2Q_2^5+8Q_3^5 \nonumber \\
&   &   -Q_1^5+Q_2^5  \ \ \  ;\ \ \  -Q_1^9+Q_2^9
\eeqa
The first two linear combinations couple to twisted RR fields with $k=1,5$
whereas the other two couple to RR fields with $k=2,4$.
Notice that these linear combinations may  be directly obtained from the
linearly independent rows of the matrices in eq.(\ref{matrices}).

The mixed $U(1)_i$ gravitational anomalies are
proportional to $(12,-12,0,12,-12,0)$. Using the sum rule
(\ref{sumrule}) it is easy to check that they are also cancelled.
The same is true for the cubic anomalies.

\medskip

{\it iv) Further compact orientifold examples }

The $U(1)$ anomalies of the compact orientifolds  discussed in refs.\cite{afiv}
all cancel in an analogous way.
The number of anomalous $U(1)$'s in each model may be found by
computing how many linearly independent generators $Q_k(\beta , p)$
in eq.(\ref{combi}) the given orientifold has.
In the case of odd order $Z_N$
orientifolds the number of anomalous $U(1)$'s is easy to guess.
There is one anomalous $U(1)$ per twisted sector $k$ (and its conjugate)
leaving no fixed tori, yielding a total of $(N-1)/2$. Thus for
 $Z_3$ and $Z_7$ we have one and three anomalous $U(1)$'s, respectively.
For $Z_3\times Z_3$  there is only one anomalous $U(1)$ since there is only
one twisted sector leaving no fixed tori. In the presence of Wilson lines
there are additional anomalous $U(1)$'s. Thus adding one Wilson line
in an odd $Z_N$ orientifold yields $N$ more $U(1)$'s, with $(N-1)$ of
them anomalous and the other decoupled from the matter fields.

In the case of even $Z_N$ orientifolds finding out the number of
anomalous $U(1)$'s depends in the detailed
structure of each twisted sector. We find three  for $Z_6$,
four for $Z_6$' and five for $Z_{12}$. The total number of $U(1)$'s
for these three cases is six, six and twelve respectively.
The other $Z_N$ cases which act crystalographically were shown to
have tadpoles (end hence to be inconsistent) in ref.\cite{afiv}  .

\section{Brane-box models and D3-branes at $Z_N$ orbifold
and orientifold singularities}

An interesting application of the formalism developed above is the study
of U(1)'s in the field theory of D3 branes sitting at orbifold and
orientifold
singularities $\IC^3/\Gamma$. There are several motivations to study such
system. First, they provide a simple construction of $D=4$, $N=1$ {\em
finite} field theories \footnote{In the case of orientifold
singularities, the field theories are finite only in the limit of large
number of D3 branes.} \cite{fin,ks,lnv,bkv,kakush}.
This has adquired certain relevance by the recent developments
of the AdS/CFT correspondence \cite{ads}, as has been analyzed in a number
of papers (following the approach of \cite{ks}). Second, the configuration
of D3 branes at {\em abelian} orbifold singularities is T-dual of the
brane box models introduced in \cite{hz}, as shown in \cite{hu}. These
brane configurations have been introduced as a tool to construct large
families of chiral $D=4$, $N=1$ gauge theories (other related
configurations have been proposed in \cite{lpt}). It is expected that
a better understanding of these brane constructions will provide insight
into the quantum effects in four-dimensional chiral gauge theories.
Finally, it is
useful to analyze the case of non-compact orbifolds/orientifolds (as
opposed to the compact toroidal orientifolds studied above) since they are
much less restricted. For instance, the orbifold group is not required
to act crystalographically. Also, some tadpole cancellation
conditions
need not be imposed, due to the non-compactness of the space.
Thus, one can study large families of examples, and the general formulae
for the cancellation of their $U(1)$ anomalies illustrate very clearly the
appearance and importance of each term in the factorization expression
(\ref{master}).

We start studying the case of D3 branes at orbifold singularities, and
then turn to a family of new $D=4$ $N=1$ gauge theories from D3 branes at
orientifold singularities.

\subsection{D3 branes at orbifold singularities}

In the following we will center on the theory of D3 branes sitting at
$\IC^3/\IZ_N$ orbifold singularities (the case of $\IC^3/(\IZ_N\times
\IZ_M)$ can be analyzed in complete analogy). These theories have been
studied in \cite{dgm,lnv}. Let the generator $\theta$
of $\IZ_N$ act on $\IC^3$ through the twist vector
$v=(\ell_1,\ell_2,\ell_3)/N$, with $\ell_1+\ell_2+\ell_3=0$. Let us also
define the action of $\theta$ on the Chan-Paton factors of the D3 branes
to be given by a diagonal matrix $\gamma_{\theta}$ which has $n_j$
eigenvalues equal to $\exp(2i\pi j/N)$, with $j=1,\ldots N$.

The gauge group of the resulting $D=4$, $N=1$ field theory on the D3
brane world-volume is
\beq
\prod_{j=1}^N U(n_j)
\label{orbigroup}
\eeq
and there are chiral multiplets transforming as
\beq
\bigoplus_{a=1}^3 \bigoplus_{j=1}^N (\fund_j,\antifund_{j+\ell_a})
\label{orbimatter}
\eeq
where the subindices denote the group under which the field transforms.
The $U(1)$ charges are normalized such that the fundamental representation
$\fund_j$ has charge $+1$ under $U(1)_j$.

As discussed in \cite{hu}, these orbifold models are T-dual to some Type
IIB configurations of NS, NS$'$ and D5 branes (brane box configurations)
introduced in \cite{hz}. It is a simple matter
to obtain a brane box model yielding the above spectrum \footnote{Other
brane box models may also give the same spectrum. All box models with the
same field theory are related by permutations in the correspondence
between the three complex coordinates in the singularity picture and the
three kinds of arrows in the brane box construction. For details, see
\cite{hu}.}. First one
constructs an infinite array of boxes using the NS and NS$'$ branes. Next,
one puts labels (ranging from $1$ to $N$, and defined modulo $N$) in the
boxes, in such a way that when one moves horizontally one box to the left
the label shifts by a quantity $\ell_1$, and when one moves vertically
one box upwards the label shifts by $\ell_2$ (this already ensures that
when one moves diagonally one box from upper right to lower left, the
label shifts by $\ell_3$). The number of D5 branes
in the box with the $i^{th}$ label is set to be $n_i$. This construction
ensures that, when one applies the rules derived in \cite{hz}, the gauge
group and matter content of the field theory obtained are as in
(\ref{orbigroup}), (\ref{orbimatter}).

Boxes with identical labels should be identified, so the array must have a
certain periodic structure. This means the corresponding two dimensional
plane is compactified to a two-torus. These are the elliptic models first
introduced in \cite{hsu}.

One of the open questions in the construction of brane box models is to
determine the restrictions on possible sets $\{ n_{i,j} \}$ of numbers of
D5 branes in each
box. From the field theory point of view, there are constraints from
cancelation of
non-abelian anomalies. From the viewpoint of consistency of the string
theory construction, they are expected to follow from conservation of RR
charge (as e.g. in the brane construction of chiral
six-dimensional theories \cite{brane6d}). However, the brane box
configurations are rather complicated, and all attempts to derive
these restrictions from string theory have failed.

The situation in the T-dual picture is better suited for a string theory
analysis. Here all NS fivebranes have been transformed into an orbifold
singularity, where a set of D3 branes sit. The configuration can be
studied in
string perturbation theory, which allows to see the one-loop effects
responsible for the anomalies. This analysis has been performed
in \cite{lr}, where it was shown that cancellation of certain tadpoles
implied the cancellation of non-abelian anomalies. Specifically, the
twisted tadpoles required to vanish correspond to twists whose only fixed
point is the origin. For twists with a two real dimensional set of fixed
points, there are non-compact dimensions along which the corresponding RR
flux can escape to infinity, and tadpoles are related to physical
quantities in the field theory, namely the one-loop beta functions of the
gauge factors.

The outcome of this analysis is that the string consistency conditions,
in the form of cancellation of mentioned tadpoles, amounts to the
rule \cite{lr}
\beq
{\rm either} \;\; n_i=0 \;\; {\rm or} \;\;
A_i\equiv \sum_{a=1}^3 (\, n_{i+\ell_a} - n_{i-\ell_a}\,)=0.
\label{conditions}
\eeq
Observe that these conditions imply the cancellation of non-abelian
anomalies, but are in fact slightly stronger. In particular, when
$n_i=1,2$, the non-abelian part of the $i^{th}$ group factor does
not exist or is $SU(2)$ and thus automatically anomaly-free. In such
case, since $n_i\neq 0$, the conditions impose $A_i=0$, even though there
is no field theory reason for it. Our analysis of $U(1)$ anomalies below
will shed some light about why the condition $A_i=0$ is required even in
those cases.

\medskip

In the following we show that the factorization of $U(1)$ anomalies
proposed in Section~2 follows from the
string consistency conditions (\ref{conditions}). We start by studying the
mixed anomaly with non-abelian factors.
For simplicity, let us first assume that all factors contain a
non-abelian part. The basic argument goes as follows. The mixed anomaly
between the $j^{th}$ $U(1)$ and the $SU(n_l)$ factor, as computed from the
spectrum (\ref{orbigroup}), (\ref{orbimatter}), is proportional to
\beqa
A_{jl} \,=\, \frac 12 n_j \sum_{a=1}^3 \big(\delta_{l,j+\ell_a}
- \delta_{l,j-\ell_a} \big) \; +
\frac 12 \delta_{j,l} \sum_{a=1}^3 \big( n_{j+\ell_a} - n_{j-\ell_a} \big)
\label{orbinonab}
\eeqa
Notice that the second contribution is proportional to the
non-abelian anomaly coefficient $A_j$ which must vanish for consistency of
the theory. The remaining contribution can be recast as
\beqa
A_{jl} \,=\, \frac {1}{2N} n_j \sum_{a=1}^3 \sum_{k=1}^N
\big\{ \exp[2i\pi (j+\ell_a-l)k/N] - \exp[2i\pi (j-\ell_a-l)/N]
\big\}
\eeqa
where we have used the discrete Fourier transform of the Kronecker delta.

Factorizing the term $\exp[2i\pi(j-l)k/N]$, and using the identity
$\sum_{a=1}^3 \sin 2\pi kv_a$ $=-4\prod_{a=1}^3 \sin \pi k v_a$, where
$v_a=\ell_a/N$, we have
\beqa
A_{jl}\, =\, \frac{(-i)}{2N} \sum_{k=1}^N  \big[\, n_j \exp(2i\pi \, kj/N)
\exp(-2i\pi \, kl/N) \prod_{a=1}^3 2\sin \pi k v_a \,\big]
\eeqa

Comparing with eq. (\ref{master}), we see it has the adequate structure to
be cancelled by the exchange of twisted RR fields (notice that, due to
the absence of orientifold projection, in the orbifold case the traces in
(\ref{master}) amount to exponentials). A nice insight our general formula
reveals is that
the coefficients $C_k$ in (\ref{generalck}), that arise from the cylinder
diagram, are directly related (through the
Fourier transform) to the delta functions which define the matter content
of the gauge theory.

It is easy to extend the conclusion to the case where some abelian or
non-abelian factor is absent. Then, the corresponding mixed anomalies
vanish, but so do the contribution from exchange of RR fields.

\medskip

The analysis of the cubic $U(1)_j\times U(1)_l^{\,2}$ anomaly works
similarly.
Let us first consider the case $n_i\neq 0$). The anomaly is proportional
to
\beqa
B_{jl} \,=\, \sum_{a=1}^3 \big( n_j\, n_{j+\ell_a}\,
\delta_{l,j+\ell_a} - n_j\, n_{j-\ell_a}\, \delta_{l,j-\ell_a}\big) \; +
\sum_{a=1}^3 \big(  n_j\, n_{j+\ell_a}\, \delta_{j,l} -
n_{j-\ell_a}\, n_j\,  \delta_{j,l} \big)
\eeqa

The second contribution is proportional to $A_j$, which
vanishes due to tadpole cancellation. The remaining part can be written
as
\beqa
B_{j,l}\, = \, n_j n_l\sum_{a=1}^3 \big( \delta_{l,j+\ell_a}
- \delta_{l,j-\ell_a}\big)
\eeqa
which factorizes as in the previous case.
Notice that, since no $\ell_a=0$ mod $N$, we have $B_{i,i}=0$ and the
subtlety about the symmetry factor $1/3$ discussed in Section~3 is not
manifest
in this family of models. Again, appropriate factorization is also found
in models where some abelian factor is absent, {\em i.e.} some $n_i=0$.

\medskip

We have shown how the string consistency conditions imply the
factorization of $U(1)$ anomalies. A nice result is that this provides a
field theory interpretation for the
requirement of vanishing of $A_i$ even when $n_i=1$. It is not required to
cancel non-abelian anomalies, but to ensure factorization of $U(1)$'s.
Another interesting outcome of our analysis is that the mecanism of
Section~5 generates masses for the $U(1)$'s. This phenomenon is
responsible for the disappearance of the $U(1)$ factors from the
low-energy dynamics, and is to be interpreted as the `freezing' described
in the language of brane construction in \cite{wit4d}. The phenomenon is
analogous to that present is six-dimensional theories from D5 branes on
singularities \cite{blumintri}.

\medskip

It is interesting that the requirement of factorization of
$U(1)$'s imposes constraints on the spectrum of the model beyond those
following from cancellation of non-abelian anomalies. These restriction
can be translated to constraints in the T-dual brane box configurations.
In order to illustrate this feature, consider a $\IC^3/\IZ_3$
singularity, at which we place D3 branes with the following Chan-Paton
matrix
\beq
\gamma_{\theta}\; =\; {\rm diag}( I_4, \alpha I_2, \alpha^2 I_2 )
\eeq
where $\alpha=e^{2i\pi/3}$ and $I_k$ is the $k\times k$ identity matrix.
Thus we have $n_0=4$, $n_1=n_2=2$. The spectrum of the models is
\beqa
& U(4)\times U(2)\times U(2) &\nonumber \\
& 3(4,2,1) + 3(1,2,2) + 3({\ov 4},1,2) &
\eeqa
This is a priori a phenomenologically interesting model since
it consists of three standard Pati-Salam
$SU(4)\times SU(2)_L\times SU(2)_R$ generations
plus three sets of Higgs fields coupling to them.
Only a set $(4,1,2)+({\overline 4},1,2 )$ is missing in order to
further do the symmetry breaking to the standard model.

This model is free of non-abelian anomalies, by virtue of the special
properties of $SU(2)$. However, it is not consistent from the string theory
point of view, since $A_1$, $A_2$ are non-vanishing. This shows up as a
non-vanishing contribution to the first term in (\ref{orbinonab}), which
spoils the factorization of $U(1)$'s.

\subsection{D3 branes at orientifold singularities}

In this subsection we study factorization of $U(1)$'s in a family of
field theories arising from D3 branes at orientifold singularities
({\em i.e.} orbifold singularities with some world-sheet orientation
reversing projection).

Such four-dimensional $N=1$ theories have only been constructed for a few
orientifold singularities \cite{kakush}. Below we construct an
infinite family of field theories corresponding to an infinite family of
orientifold singularities. These field theories are very interesting,
since by arguments from the AdS/CFT correspondence,
they will be superconformal in the large $N$ limit. Moreover, they
constitute the first example of an infinite family of $N=1$ theories from
D3 branes at orientifold singularities, and show that D3 branes at
orientifold singularities generate a rich variety of $D=4$, $N=1$ field
theories. We leave a more systematic study of other possible families of
models for future work, and in what follows center on a particular large
class. The details of the
tadpole computation showing its consistency are left for the appendix.
Here
we just state the main properties of the spectrum.

We will consider a discrete group $\IZ_N$, with odd $N=2P+1$, generated
by a twist $\theta$ acting on $\IC^3$ as defined by $v=(\ell_1,\ell_2,\ell_3)$.
The orientifold group will be generated by $\theta$ and
$\Omega'\equiv\Omega
(-1)^{F_L}R_1 R_2 R_3$, where $R_a$ denotes the inversion of the $a^{th}$
complex plane. As usual, $\Omega$ includes the element $J$ which exchanges
oppositely twisted sectors. Notice that this orientifold does not require
the presence of D7 branes.

Before the $\IZ_2$ orientifold projection, the spectrum of the model is
\beqa
\prod_{j=-P}^P U(n_j)\quad ; \quad \bigoplus_{a=1}^3 \bigoplus_{j=-P}^P
(\fund_j, \antifund_{j+l_a})
\label{orient}
\eeqa
just like in (\ref{orbigroup}), (\ref{orbimatter}). The effect of
$\Omega'$ is to exchange the factors $U(n_j)$ and $U(n_{-j})$, in such a
way that the fundamental representation $\fund_j$ goes over to the
anti-fundamental representation $\antifund_j$, and vice-versa. Notice
that, in order to be a symmetry of the theory (\ref{orient}), we must
require the ranks of exchanged groups to be equal, $n_j=n_{-j}$.
The operation is an automorphism of the quiver diagram of the theory.

The final spectrum is found by keeping the fields invariant under the
orientifold projection. The absence of D7 branes in the construction
allows for two different projections. In one of them, the group $U(n_0)$,
which is projected onto itself, becomes $SO(n_0)$ in the
quotient. Also, when the two entries of some bi-fundamental are charged
with respect to the same group in the quotient, the antisymmetric
combination is to be taken. The second possibility is to project onto
$USp(n_0)$, and symmetric representations. In the following, we will
center on the `$SO$' projection, and we stress the other case works
in complete analogy.

\medskip

Just to give a flavour of the type of theories that arise from the above
construction, we give a simple example of a non-crystallographic case
which has not been considered in the literature. It is a $\IZ_5$ model,
generated by a twist $v=(1,1,-2)/5$. The spectrum (\ref{orient}) before
the orientifold projection is
\beqa
& U(n_{-2}) \times U(n_{-1}) \times U(n_{0}) \times U(n_{1}) \times
U(n_{2}) & \nonumber\\
& 2\; \big[\, (\fund_1,\antifund_2) + (\fund_2,\antifund_{-2}) +
(\fund_{-2},\antifund_{-1}) + (\fund_{-1},\antifund_{0}) +
(\fund_{0},\antifund_{1})\, \big] + & \nonumber \\
& + (\fund_1,\antifund_{-1}) + (\fund_{2},\antifund_{0}) +
(\fund_{-2},\antifund_{1}) + (\fund_{-1},\antifund_{2}) +
(\fund_{0},\antifund_{-2}) &
\eeqa
After the orientifold projection, the rules above yield the spectrum
\beqa
& SO(n_0) \times U(n_1)\times U(n_2) & \nonumber \\
& 2 \;\big[\, (\fund_1,\antifund_2) + \Yasymm_2 + (\fund_0,\antifund_1)\,
\big] + & \nonumber \\
& + \Yasymm_1 + (\fund_0,\fund_2) + (\antifund_1,\antifund_2) &
\label{specz5}
\eeqa

The particular cases of $\IZ_3$ and $\IZ_7$ have appeared in
\cite{kakush}, and, in the T-dual version of D9 branes in a compact
orientifold, in \cite{ang,afiv,kak1,kak2}.

\medskip

Since this family of theories has not appeared in the literature, we make
here a brief comment concerning their non-abelian anomalies.
The non-abelian anomaly coefficient, in the generic case in which no $n_i$
vanishes, is given by
\beq
A_i \; =\; \sum_{a=1}^3 \big[\, (n_{i+\ell_a}-n_{i-\ell_a}) \, -\,
4(\delta_{i+\ell_a,-i}-\delta_{i-\ell_a,-i}) \,\big]
\label{orientai}
\eeq
where the second contribution takes into account the cases where the
bifundamental is actually an antisymmetric. In this and following
expressions, the formulae are valid for the `Sp' projection by simply
changing the sign of such contributions.

In analogy with the orbifold case \cite{lr}, the conditions $A_i=0$ can be
stated as constraints on the Chan-Paton matrices, simply by taking
a discrete Fourier transform. After a short computation, the conditions
$A_i=0$ are equivalent to
\beq
\prod_{a=1}^3 \sin 2\pi k v_{a} \; \Tr \gamma_{2k} \,-\, 4 \prod_{a=1}^3
\sin \pi k v_a =0
\eeq
or
\beq
\prod_{a=1}^3 \sin \pi k v_a  \left[ 2 \prod_{a=1}^3 \cos \pi k v_a \Tr
\gamma_{2k} - 1 \right]=0
\eeq

The results of the appendix show that string consistency conditions
actually ensure the vanishing of this quantity. The first factor is
non-zero when the twist has the origin as the only fixed point. This is
precisely when tadpoles, which are proportional to the second factor,
are required to vanish. This shows how string consistency implies the
consistency of the gauge field theory on the D3 branes. Notice that in 
\cite{kakush} only the $\IZ_3$ and $\IZ_7$ models were
considered, since the indirect contruction technique employed there (the
system of D3 branes was obtained by T dualizing  a set of D9 branes) does
not allow to obtain the whole infinite family.

Being a bit more careful, it is possible to show that if some $n_i$
vanishes the condition for $A_i$ is not required. Notice that, as in the
orbifold case, when some $n_i=1,2$ the condition $A_i$ does not have the
interpretation of cancellation of non-abelian anomalies. Our arguments
below will show that it is however needed to have appropriate factorization of
$U(1)$ anomalies.

\medskip

So, let us center on the study of mixed non-abelian anomalies. We first
consider the case where no non-abelian factor is absent.
The mixed anomaly between the $j^{th}$ $U(1)$ and the $l^{th}$ non-abelian
factor can be computed from the spectrum of the theory to be
\beqa
A_{jl} & = & \frac 12 \delta_{j,l} \sum_{a=1}^3 \big(n_{l+\ell_a} -
n_{l-\ell_a} \big)
+ \frac 12 n_j \sum_{a=1}^3 \big[ \big( \delta_{l,j+\ell_a} -
\delta_{l,j-\ell_a} \big) + \big( \delta_{l,-j-\ell_a} -
\delta_{l,-j+\ell_a} \big) \big] - \nonumber\\
& & -2 \sum_{a=1}^3 \delta_{j,l} \big( \delta_{j+\ell_a,-j}+
\delta_{j-\ell_a,-j} \big)
\eeqa

The main difference with the equivalent expression for the orbifold case
is that, besides the contribution from matter charged under the $j^{th}$
and $l^{th}$ factors, there is an additional contribution from matter
charged under the $j^{th}$ and $(-l)^{th}$ factors. This last contribution
arises in string theory from the second diagram in Figure~1, the
Moebius strip. The last line contains a correction that takes
into account the cases when some bi-fundamental is actually an
antisymmetric (recall that in our normalization, the charge of an
antisymmetric representation is $+2$).

The term proportional to $\delta_{j,l}$ in the above expression is
proportional to $A_l$, eq.(\ref{orientai}), and must vanish for string
consistency. The remaining terms can be Fourier transformed in a by now
familiar fashion, to yield
\beq
A_{j,l}\; =\; 1/N \sum_{k=-P}^P n_j \sin (2\pi kj/N) \cos (2\pi kl/N)
\prod_{a=1}^3 2\sin \pi k v_a
\eeq

This reproduces the structure depicted in (\ref{master}), or
(\ref{masterorient}). Also, it nicely shows how the orientifold projection
implies the appearance of sine and cosine functions instead of
exponentials.

Again, when some abelian or non-abelian factor is absent, the
corresponding $A_{j,l}$ vanishes automatically, but so does the exchange
of RR fields.
\medskip

Let us finally discuss the structure of cubic anomalies. Since
factorization works analogously, the discussion will be brief. We
center on the generic case of $n_i\neq 0$. The mixed
$U(1)_j\times U(1)_l^2$ anomaly is given by
\beqa
B_{j,l} = & \delta_{j,l} \sum_{a=1}^3 \big(n_j\, n_{l+\ell_a} -
n_{l-\ell_a} n_j \big)
+ n_j\, n_l\, \sum_{a=1}^3 \big[\, \big( \delta_{l,j+\ell_a} -
\delta_{l,j-\ell_a} \big) + \big( \delta_{l,-j-\ell_a} -
\delta_{l,-j+\ell_a} \big)\, \big] - \nonumber\\
 & -\delta_{j,l} \sum_{a=1}^3 \big[\, (2 n_j^2-4n_i) (\delta_{j+\ell_a,-j}
- \delta_{j-\ell_a,-j})\, \big]
\eeqa

As usual, some terms can be grouped to yield a contribution proportional
to $A_j$, eq. (\ref{orientai}), which vanishes. The remaining terms
are
\beqa
B_{j,l} & =  n_j\, n_l \sum_{a=1}^3 & \big[\, \delta_{l,j+\ell_a} +
\delta_{l,-j-\ell_a} - \delta_{l,j-\ell_a} -\delta_{l,-j+\ell_a} + \nonumber \\
& & +2\,\delta_{j,l}\, \big(\delta_{j,-j-\ell_a} - \delta_{j,-j+\ell_a} +
\delta_{j,j+\ell_a}-\delta_{j,j-\ell_a} \big)\,\big]
\eeqa
The last two terms in the second line, which are vanishing because
$\ell_a\neq 0 \;{\rm mod} \; N$, and so $j\neq j\pm \ell_a$, have been
introduced for convenience.

When $i\neq j$, the second line contribution vanishes, and the expression
factorizes in the same fashion as the mixed non-abelian anomalies (with
the additional normalization factor $n_l$). When
$i=j$, the contribution from the second line has the same form as that
from the first, and both together give the usual factorized form with an
additional factor of $3$, which cancels agains the symmetry factor $1/3$
mentioned in Section~3.

\medskip

We have shown how this large family of field theories constructed from D3
branes at orientifold singularities satisfy in a very non-trivial way all
the constraints of cancellation of non-abelian anomalies, and appropriate
factorization of $U(1)$ anomalies. These properties follow directly from
the cancellation of non-physical tadpoles (\ref{tadpole}). This family
illustrates the rich variety of field theories from D3 branes at
orientifold singularities, and hopefully will motivate further research
in the field.

\section{Fayet-Illiopoulos terms}

The anomaly cancellation mechanism described in previous sections relies
on the
presence of the couplings (\ref{contrat}) which mix the $U(1)$ fields with the
RR two-forms $B_k$.
In addition, supersymmetry requires the presence of  terms of the
form
\beq
D_i\ \sum_{k=1}^N (\Tr \gamma_k\lambda _i)\ \Phi _k
\label{fayet}
\eeq
where $D_i$ is the auxiliary field asociated to the $U(1)_i$ and
$\Phi _k$ is the NS partner of the corresponding RR fields $\phi _k$.
These are nothing but field-dependent Fayet-Illiopoulos terms.
They are similar to the FI terms found for $D=6, N=1$ in \cite{dm}.
Notice that in the case of orientifolds one has
\beq
D_i \ \sum _{k=1}^N (n_i \sin 2\pi kV_i )\ \Phi _k
\label{fayeto}
\eeq
and untwisted $(k=0)$ NS fields (like e.g., the dilaton or untwisted moduli) do not
contribute to FI-terms. This is quite different to the
case of $N=1, D=4$ heterotic vacua in which it is only the dilaton
which appears in the FI term \cite{anouno,fi} .

Other differences with the case of $D=4, N=1$ heterotic vacua are worth remarking.
In the heterotic case
there is at most one anomalous $U(1)$ whereas in the
Type I and Type II vacua studied in this paper any number may appear.
Furthermore, the counterterm (\ref{contrat})  appears here at the disk
level whereas in the heterotic such term is induced at the one-loop level.

Moreover, in the heterotic case, since the FI term is proportional to the
heterotic dilaton, one
cannot put it to zero without going to a non-interacting $(g=0)$ theory.
Thus the scale of the heterotic FI term is of order of the string scale or
slightly below.

In the Type I or Type II cases here considered we can in principle
put the size of the FI-terms as small as we wish since $\Phi_k \rightarrow
0$
does not correspond in general to a non-interacting theory,
but to the orbifold limit. Consider for
example the case of the compact orientifolds discussed in chapter two.
The gauge kinetic function for a $U(1)_j$ will have the
general form
\beq
f_j\ =\ S\ +\ \sum _{k} n_jcos2\pi kV_j \ \Psi _k
\label{ffunction}
\eeq
where $S$ is the Type I untwisted dilaton chiral field and
$\Psi _k =\Phi _k +i \phi _k $ are the complex twisted scalars.
For $\Phi _k \rightarrow 0$ there is a finite coupling
given by the untwisted dilaton.
The same is true for the kinetic functions of non-Abelian
group factors for which one can write a similar expression.
Notice  that, since the cosine in eq.(\ref{ffunction})
may have both positive and negative sign, for some particular points
in the dilaton/twisted NS fields moduli space, some of the gauge
coupling constants may explode. At those points perturbation theory will fail
and one expects the appearence of non-perturbative phenomena.

 The couplings in eq.(\ref{contrat}) (after a duality transformation)
imply the presence of a
Higgs mechanism by which $U(1)$'s get masses. The masses of these
$U(1)$'s , like the FI-terms, are given by the vev's of the
NS $\Phi _k$ fields. Thus the masses of the anomalous $U(1)$'s
in these theories may in principle be as small as we wish.
This is again to be contrasted to the heterotic case
in which the mass of the unique anomalous $U(1)$ is tied up to
the string scale.

\section{Final comments}

In this paper we have shown that four-dimensional type IIB orientifold
vacua have generically several anomalous $U(1)$'s. We have discussed how
their triangle anomalies are cancelled through the exchange of twisted RR
fields, in a version of the Green-Schwarz mechanism. This pattern is very
different from that found in heterotic models, as we have already
remarked, and seems to be worth of further study.

Here would like to stress another point concerning this mechanism.
Four-dimensional $N=1$ vacua can also be obtained by compactifying type I
superstrings on smooth Calabi-Yau manifolds with a certain gauge bundle.
These models can be analyzed as a Kaluza-Klein reduction of the
ten-dimensional theory, and yield models with a pattern of $U(1)$
anomalies identical to that in heterotic models. Namely, there is at most
{\em one} anomalous $U(1)$, and the anomaly is cancelled by exchange of
the partner of the dilaton \cite{wit32}. This is somewhat surprising,
since the
orientifold models we have analyzed (at least, those having no D5 branes)
can be naively regarded as singular limits of such smooth
compactifications. However, both constructions differ sharply in their
pattern of anomaly cancellation.

Nevertheless, this is not the first time that such differences between
compactifications on smooth and singular manifolds are found.
Already in six-dimensional $N=1$ vacua, smooth compactifications of type I
string theory yield models with at most one tensor multiplet, containing
the dilaton. On the other hand, orientifold models generically contain
additional tensor multiplets, arising from the singular points
\cite{gj,dp2}. Understanding the relation between these two kinds of vacua
has been the key to some of the new physics uncovered in six-dimensional
string and field theory.

The situation we encounter in $D=4$, $N=1$ vacua is certainly analogous.
Four-dimensional orientifolds seem to be exploring regions of the
moduli space which differ in their generic properties from those of smooth
compactifications. One is led to expect new interesting insights on $D=4$,
$N=1$ vacua will be obtained by studying the relation between both
descriptions.

On the phenomenological side, the anomalous $U(1)$'s found
in the class of orientifold/orbifold Type IIB vacua  studied
in the present paper show characteristics totally different
to the familiar single $U(1)$'s of perturbative heterotic
vacua. It would be interesting to study in more detail
possible phenomenological applications of these new
anomalous $U(1)$'s to problems like fermion textures,
supersymmetry-breaking and cosmology in which the
heterotic anomalous $U(1)$'s have been suggested to
play an important role.

\bigskip

\bigskip

\bigskip

\centerline{\bf Acknowledgements}

A.M.U. thanks J.~Park for many useful discussions, and
M.~Gonz\'alez for encouragement and support.
L.E.I. thanks G. Aldazabal, A.~Font and G. Violero for
sharing with him many of their insights in
orientifold construction.
L.E.I. and R.R. thank CICYT (Spain)
and IBERDROLA  for financial support.
The work of A.M.U. is supported by the Ram\'on Areces Foundation (Spain).
L.E.I. thanks CERN Theory Division where part of this work
was carried out.

\newpage

\section{Appendix}

{\large \bf Construction of the theories on D3 branes at orientifolds}

In this appendix we construct the non-compact orientifold theories of
Section~4.2 and discuss their tadpole cancellation conditions. Several
remarks were already made in the main text, but we repeat them here for
convenience.

The orientifold group has the structure $G_{orient}=G+\Omega'G$. Here $G$
is a $\IZ_{N}$ group with $N=2P+1$, odd, generated by a twist $\theta$
with vector $v=(v_1,v_2,v_3)$$={1\over N}(\ell_1,\ell_2,\ell_3)$. The
world-sheet orientation reversing operation is $\Omega'\equiv \Omega
(-1)^{F_L}R_1 R_2 R_3$, where the operator $R_a$ is the inversion of the
$a^{th}$ complex plane. As usual $\Omega$ exchanges oppositely twisted
sectors (it is $\Omega J$ in the notation of \cite{polten} ).

The orientifolds do not contain D7 branes, and thus there are two
possible $\Omega'$ projections on the D3 branes, which we will denote as
the `$SO$' and `$Sp$' projections.
Notice also that these models are four-dimensional cousins of some
$D=6$ orientifolds considered in \cite{gj,dp2}.

The Chan-Paton matrices we will consider are
$$\gamma_{\theta}\ ={\rm diag}(I_{n_0},\alpha I_{n_1},\ldots,\alpha^P
I_{n_P},\alpha^{-P} I_{n_P},\ldots,\alpha^{-1} I_{n_1}) \quad
{\rm with}\;\; \alpha=e^{2\pi i/N}
$$
{\scriptsize
$$\gamma_{\Omega'}\ =\ \pmatrix{
I_{n_0} &  &  &   &   &  & \cr
&  &  &   &   &   & I_{n_1} \cr
&  &  &   &   & \cdots  &   \cr
&  &  &  &  I_{n_P} &   & \cr
&  &  &  I_{n_P} &   & & \cr
&  & \cdots &   &   &   & \cr
& I_{n_1}  &  &   &   & \cr
} \quad ;\quad
\gamma_{\Omega'}\ =\ \pmatrix{
\epsilon_{n_0} &  &  &   &   &  & \cr
&  &  &   &   &   & I_{n_1} \cr
&  &  &   &   & \cdots  &   \cr
&  &  &  &  I_{n_P} &   & \cr
&  &  &  -I_{n_P} &   & & \cr
&  & \cdots &   &   &   & \cr
& -I_{n_1}  &  &   &   & \cr
}
$$}
where $\epsilon_{n_0}$ is block-diagonal with $n_0/2$ blocks of the form
{\scriptsize $\pmatrix{0 & 1 \cr -1 & 0 }$}.
The two possibilities for $\gamma_{\Omega'}$ correspond to the
$SO$ and $Sp$ projections, respectively.
For future convenience, notice that the matrices verify
\beq
\Tr(\gamma_{\Omega'_k}^{-1} \gamma_{\Omega'_k}^T)=\pm \Tr(\gamma_{2k})
\label{property}
\eeq
where the upper (lower) sign corresponds to the $SO$
($Sp$) projection.

The spectrum before the $\Omega'$ projection is given by the orbifold
theory
\beqa
\prod_{i=-P}^P U(n_i) \quad ; \quad
\bigoplus_{a=1}^3 \bigoplus_{i=-P}^P (\fund_i,\antifund_{i+l_a})
\eeqa

The operation $\Omega'$ identifies the groups $U(n_i)$ and $U(n_{-i})$,
such that the representation $\fund_i$ is identified with
$\antifund_{-i}$. In the $SO$ (resp. $Sp$)projection, $U(n_0)$ becomes
$SO(n_0)$ (resp. $USp(n_0)$), and bi-fundamentals charged with respect to
the same group in the quotient become antisymmetric (resp. symmetric)
representations.

{\bf Computation of tadpoles}

Since we are quotienting $\IC^3$ and the resulting space is non-compact,
string consistency only requires the cancellation of tadpoles for twisted
sectors which have the origin as the only fixed point \footnote{However,
in order to have a conformal theory in the limit of large number of D3
branes, all twisted tadpoles should vanish (at leading order)}. We
mainly refer to the general expressions
derived in the appendix of \cite{afiv}, and only mention what changes
should be taken into account.

The 33 cylinder amplitudes is identical to that of 99 or 55 cylinders.
>From equations (7.9) and (7.14) in \cite{afiv}, we have the $t\to 0$
contribution
\beqa
\cc_{33} \rightarrow (1-1)\frac{V_4}{8N} \sum_{k=1}^{N} \int_{0}^{\infty}
\frac{dt}{t} (8\pi \alpha' t)^{-2} t \prod_{a=1}^{3} |\, 2\sin \pi kv_a
\, |  (\Tr \gamma_{k,3})^2
\eeqa
For the Klein bottle, there is only the contribution
$\cz_{\ck}(1,\theta^k R_1R_2R_3)$, where we must also include the
twists
implicit in $\Omega'$. From equations (7.1) and (7.6) in \cite{afiv}, we
have the $t\to 0$ contribution
\beqa
\ck \rightarrow (1-1)\frac{V_4}{8N} \sum_{k=1}^{N} \int_{0}^{\infty}
\frac{dt}{t} (4\pi \alpha' t)^{-2} (2t) \frac{\prod_{a=1}^{3}
|\,2\sin[2\pi (kv_a+1/2)]\,|}{\prod_{a=1}^3 4 \sin^2[\pi(kv_a+1/2)]}
\eeqa
The shifts by $1/2$ arise from the twist $R_1R_2R_3$, and the denominator
is the zero mode integration mentioned in \cite{gj} (in the
compact examples in \cite{afiv} it was taken into account by an explicit
counting of fixed point sets).

Finally, the contribution from the Moebius strip $\cz_3(\theta^k
R_1R_2R_3)$ is analogous to eq.(7.24) in \cite{afiv}, but for the fact
that there are Dirichlet boundary conditions all three complex planes.
Using also (7.20) in \cite{afiv}, the leading contribution as $t\to 0$ is
\beqa
\cm_{3} \rightarrow (1-1) \frac{V_4}{8N} \sum_{k=1}^{N}
\int_{0}^{\infty}
\frac{dt}{t} (8\pi \alpha' t)^{-2} t \prod_{a=1}^{3}
 s_a 2 \cos[\pi (kv_a+1/2)] \,
\Tr (\gamma_{\Omega'_k}^{-1} \gamma_{\Omega'_k}^T)
\eeqa

where $s={\rm sign} (\sin 2\pi k v_a)$.
Using the relations $t={1\over 2\ell}$, $t={1\over 4\ell}$, $t={1\over
8\ell}$ for the cylinder, Klein bottle and Moebius strip \cite{gp}, and
the property (\ref{property}), the amplitude is proportional to
\beqa
\sum_{k=1}^{N} \; \left[ \prod_{a=1}^3 |\, 2\sin 2\pi kv_a|\,  (\Tr
\gamma_{2k})^2 \mp \prod_{a=1}^3 s_a 2\sin \pi kv_a \Tr \gamma_{2k}
+ 16 \prod_{a=1}^3 \left|\,\frac{\sin \pi k v_a}{\cos \pi kv_a}\right| \;
\right]
\eeqa
with the upper (lower) sign for the $SO$ ($Sp$) projection. This can be
factorized as
\beqa
\sum_{k=1}^N \frac{1}{\prod_{a=1}^3 |\, 2 \sin 2\pi kv_a|} \left[\,
\prod_{a=1}^3 2\sin 2\pi kv_a \Tr \gamma_{2k} \mp 32 \prod_{a=1}^3
\sin \pi kv_a\,\right]^2
\label{tadpole}
\eeqa
Each of the terms in the square bracket must vanish independently. Notice
that when the twist leaves some complex plane fixed ({\em i.e.} some
$\sin \pi kv_a$ vanishes) the equation is automatically satisfied and the
tadpole corresponding to that twist does not constrain the Chan-Paton
matrices. For twists which have the origin as the only fixed point, the
constraint reads
\beqa
\Tr \gamma_{2k}= \pm \frac 12 \frac{1}{\prod_{a=1}^3 \cos \pi kv_a}
\label{tadpole2}
\eeqa

In the main text it is shown that these consistency conditions ensure the
the consistency of the field theory on the D3 branes. Namely, it implies
cancellation of non-abelian anomalies, and appropriate factorization of
$U(1)$ anomalies.

Notice that, even though in some cases these orientifolds are T-dual to
models with only D9 branes, the general formula for the tadpoles
conditions is quite different. This is always the case when the
T-duality is performed along twisted coordinates. However, the result is
consistent and for $\IZ_3$ and $\IZ_7$ the usual twisted tadpole
conditions are recovered.

A last important point is that it is always possible to find solutions to
the above tadpole equations. To show the existence of at least one for
each $Z_{2P+1}$ singularity, we rewrite the condition (\ref{tadpole2}) as
\beq
\Tr \gamma_{2k} = \pm 4 \prod_{a=1}^3 \frac{1}{1+e^{2\pi i kv_a}}
\label{soluc}
\eeq
Since $N$ is odd, the `$1$' in the numerator can be expressed as
polynomial in $\exp(2\pi i kv_a)$ with an even number of non-zero terms.
The denominator $(1+e^{2\pi ikv_a})$ divides such polynomial, and thus
$\Tr \gamma_{2k}$ is a polynomial in $\exp(2\pi k/N)$. It is then
straightforward to give $\gamma$ matrices with the appropriate traces.
This procedure gives a solution that cancels all twisted tadpoles.

As a simple example, consider the $\IZ_5$ model with $v={1\over
5}(1,1,-2)$, whose spectrum is depicted in eq.(\ref{specz5}).
From
(\ref{soluc}), we have
\beq
\Tr \gamma_{2k} =  4 \left( \frac{1}{1+\alpha^k} \right)^2
\frac{1}{1+\alpha^{3k}}
\eeq
with $\alpha=e^{2\pi i/5}$. Performing the trick mentioned above, the
fractions can be expressed as a polynomial in $\alpha$, yielding
\beqa
\Tr \gamma_{2k}= 4(-\alpha^k -\alpha^{3k})^2 (-\alpha^{3k} -\alpha^{9k})=
-4(1+\alpha^{2k}+\alpha^{2k})
\eeqa
The conditions for the twisted tadpoles are solved by
\beq
\gamma_1={\rm diag} \,(I_{N-4},\alpha I_{N-4},\alpha^2 I_N, \alpha^3 I_N,
\alpha^4
I_{N-4})
\eeq
Using the spectrum in (\ref{specz5}), one can check directly that
non-abelian
anomalies cancel.


\bigskip


\newpage

\begin{thebibliography}{99}
%
\bibitem{gs}
M. Green and J. Schwarz, \PLB{149} {84}{117}.
%
\bibitem{anouno}
M. Dine, N. Seiberg and E. Witten, \NPB{289}{87}{585} .
%
\bibitem{nignos}
J.Casas, E. Katehou and C. Mu\~noz ,
\NPB{317} {89} {171} .
%
%
\bibitem{sin}
L.E. Ib\'a\~nez, \PLB{303}{93}{55} .
%
\bibitem{fi}
J. Atick, L. Dixon and A. Sen \NPB{292}{87}{109}, \\
M. Dine, I. Ichinoise and N. Seiberg, \NPB{293}{87}{253}.
%
\bibitem{ir}
L.E. Ib\'a\~nez and G.G. Ross, \PLB{332}{94}{100}
%
\bibitem{ramond}
 For a review and references see P. Ramond, hep-ph/9604251 .
%
\bibitem{mr}
J. March-Russell, hep-ph/9806426 .
%
\bibitem{radu}
P. Binetruy, C. Deffayet, E. Dudas and P. Ramond, hep-th/9807079 .
%
\bibitem{lpst}
Z. Lalak, S. Pokorski and S. Thomas, hep-ph/9807503 .
%
\bibitem{bl}
M.~Berkooz and R.~G.~Leigh, \NPB{483} {97} {187}, hep-th/9605049.
%
\bibitem{ang}
C.~Angelantonj, M.~Bianchi, G.~Pradisi, A.~Sagnotti and Ya.S.~Stanev,
\PLB{385} {96} {96}, hep-th/9606169.
%
\bibitem{kak1}
Z.~Kakushadze, \NPB {512} {98} 221, hep-th/9704059.
%
\bibitem{kak2}
Z.~Kakushadze and G.~Shiu, \PRD{56} {97} {3686}, hep-th/9705163.
%
\bibitem{kak3}
Z.~Kakushadze and G.~Shiu, hep-th/9706051.
%
\bibitem{zwart}
G.~Zwart, hep-th/9708040.
%
\bibitem{odri}
D.~O'Driscoll, hep-th/9801114.
%
\bibitem{afiv}
G.~Aldazabal, A.~Font, L.~E.~Ib\'a\~nez, G.~Violero, hep-th/9804026.
%
\bibitem{lapolla}
Z. Kakushadze, G. Shiu and S.H. Tye , hep-th/9804092
%
\bibitem{kakn}
Z. Kakushadze, hep-th/9804110 ; hep-th/9806044 .
%
\bibitem{lykken}
J. Lykken, E. Poppitz and S. Trivedi,
hep-th/9806080 .
%
\bibitem{gepner}
R. Blumenhagen and A. Wisskirchen, hep-th/9806131 .
%
\bibitem{sagcan}
A.~Sagnotti, \PLB{294} {92} {196}, hep-th/9210127.
%
\bibitem{dm}
M.~R.~Douglas, G.~Moore, hep-th/9603167.
%
\bibitem{intri}
K.~Intrilligator, \NPB {496} {97} {177}, hep-th/9702038.
%
\bibitem{blumintri}
J.~Blum and K.~Intrilligator, Nucl.Phys. B506(1997)223, hep-th/9705030;
\NPB{506} {97} 199, hep-th/9705044.
%
\bibitem{gsw}
M.~B.~Green, J.~H.~Schwarz, P.~C.~West, Nucl.Phys. B254(1985)327.
%
\bibitem{blpssw}
M.~Berkooz, R.~G.~Leigh, J.~Polchinski, J.~H.~Schwarz, N.~Seiberg,
E.~Witten, Nucl.Phys.B475(1996)115, hep-th/9605184
%
\bibitem{hz}
A.~Hanany, A.~Zaffaroni, JHEP 05(1998)001, hep-th/9801134.
%
\bibitem{hu}
A.~Hanany, A.~M.~Uranga, JHEP 05(1998)013, hep-th/9805139.
%
\bibitem{dgm}
M.~R.~Douglas, B.~R.~Greene, D.~R.~Morrison, Nucl.Phys.B506 (1997)84,
hep-th/9704151.
%
\bibitem{lnv}
A.~Lawrence, N.~Nekrasov, C.~Vafa, hep-th/9803015,
%
\bibitem{ads}
J.~Maldacena, hep-th/9711200; E.~Witten, hep-th/9802150;
S.~S.~Gubser, I.~R.~Klebanov, A.~M.~Polyakov, Phys.Lett.B428(1998)105,
hep-th/9802109.
%
\bibitem{ks}
S.~Kachru, E.~Silverstein, Phys.Rev.Lett. 80(1998):4855, hep-th/9802183.
%
\bibitem{bkv}
M.~Bershadsky, Z.~Kakushadze, C.~Vafa, Nucl.Phys. B523(1998)59,
hep-th/9803076.
%
\bibitem{wit4d}
E.~Witten, Nucl.Phys.B500(1997)3, hep-th/9703166.
%
%
\bibitem{sagnotti}
A.~Sagnotti, in Cargese 87,
{\it Strings  on Orbifolds},
ed. G. Mack et al. (Pergamon Press, 1988) p. 521.
%
\bibitem{hor}
P.~Horava, \NPB{327} {89} {461}; \PLB{231} {89} {251};\\
J.~Dai, R.~Leigh and J.~Polchinski, Mod.Phys.Lett. A4 (1989) 2073;\\
R.~Leigh, Mod.Phys.Lett. A4 (1989) 2767.
%
\bibitem{bs} G.~Pradisi and A.~Sagnotti, \PLB{216} {89} {59};\\
M.~Bianchi and A.~Sagnotti, \PLB{247} {90} {517}.
%

\bibitem{dhvw}
L.~Dixon, J.A.~Harvey, C.~Vafa and E.~Witten, \NPB{274} {86} {285}.
%
\bibitem{fiq}
A.~Font, L.E.~Ib\'a\~nez, F.~Quevedo, \PLB{217}{89}{272}.
%
\bibitem{fin}
L.E.~Ib\'a\~nez, hep-th/9802103.
%
\bibitem{kakush}
Z.~Kakushadze, hep-th/9803214; Z.~Kakushadze, hep-th/9804184
%
\bibitem{lpt}
J.~Lykken, E.~Poppitz, S.~P.~Trivedi, Phys.Lett. B416(1998)286,
hep-th/9708134; Nucl.Phys. B520(1998)51, hep-th/9712193.
%
\bibitem{hsu}
A.~Hanany, M.~J.~Strassler, A.~M.~Uranga JHEP 06(1998)011, hep-th/9803086.
%
\bibitem{brane6d}
I.~Brunner, A.~Karch, Phys.Lett.B409(1997)109, hep-th/9705022;
JHEP 03(1998)003, hep-th/9712143. A.~Hanany, A.~Zaffaroni, hep-th/9712145.
%
\bibitem{lr}
R.~G.~Leigh, M.~Rozali, hep-th/9807082.
%
\bibitem{polten}
J.~Polchinski, \PRD {55} {97} {6423}, hep-th/9606165.
%
%
\bibitem{gj}
E.~Gimon and C.~Johnson,
\NPB{477}{96}{715}, hep-th/9604129.
%
%
\bibitem{dp2}
A.~Dabholkar and J.~Park, \NPB{477}{96}{701}, hep-th/9604178;
 \PLB{394}{97}{302}, hep-th/9607041.
%
\bibitem{gp}
E.~Gimon and J.~Polchinski,
Phys.Rev. D54 (1996) 1667, hep-th/9601038.
%
\bibitem{wit32}
E.~Witten, Phys.Lett. 149B(1984)351
%
\end{thebibliography}
\end{document}